\documentclass[conference]{IEEEtran}
\IEEEoverridecommandlockouts
\usepackage{cite}
\usepackage{amsmath,amssymb,amsfonts}
\usepackage{algorithmic}
\usepackage{graphicx}
\usepackage{textcomp}
\usepackage{xcolor}
\usepackage{booktabs}
\usepackage{diagbox}
\usepackage{comment}
\usepackage[hidelinks]{hyperref}

\newcommand{\blue}[1]{\textcolor{blue}{#1}}

\newcommand{\vqc}{\mathrm{VQC}}
\newcommand{\Emb}{\mathrm{Emb}}
\def\BibTeX{{\rm B\kern-.05em{\sc i\kern-.025em b}\kern-.08em
    T\kern-.1667em\lower.7ex\hbox{E}\kern-.125emX}}

\begin{document}

\title{What Counts as Real? Speech Restoration and Voice Quality Conversion Pose New Challenges to Deepfake Detection}


\author{\IEEEauthorblockN{Shree Harsha Bokkahalli Satish, Harm Lameris, Joakim Gustafson, and Éva Székely}
\IEEEauthorblockA{\textit{Department of Speech, Music and Hearing}
\textit{KTH Royal Institute of Technology}
Stockholm, Sweden \\
\{shbs, lameris, jkgu, szekely\}@kth.se}}

\maketitle

\begin{abstract}
Audio anti-spoofing systems are typically trained to assign one authenticity label to an entire speech utterance. This formulation becomes under-specified for transformations where the underlying speaker identity and linguistic content remain unchanged. We study this problem using benign, authenticity-preserving speech transformations, including voice quality conversion and speech restoration, applied to both bona fide and spoofed speech. Instead of treating all processed audio as spoofed, we factorise labels into source authenticity and processed status. Across SSL representations and DF-Arena fine-tuning experiments, we find that utterance processing status can transfer more reliably than source attribution: detectors can often identify that speech has been processed, while still confusing processed bona fide and processed spoofed speech. 
These results suggest that audio deepfake defences must move beyond the binary spoofed/authentic paradigm. Robust detection requires granular reporting on source authenticity, processing status, and precise processing localisation.
\end{abstract}


\section{Introduction}
The proliferation of highly realistic synthesised speech has necessitated robust countermeasures, and yet the detection landscape remains an ongoing adversarial cycle~\cite{kamel2025survey, li2024audio, zhang2025audio}. Attackers increasingly employ post-processing transformations, such as replay attacks, resulting in signals that are significantly harder to detect~\cite{muller2025replay}. In response, modern audio spoofing evaluations have shifted toward real-world variability, as seen in recent challenges~\cite{wang2024asvspoof5} and comprehensive benchmarking platforms like DF-Arena~\cite{dowerah2026speech}. As the field moves beyond simple binary detection towards the attribution of specific sources of spoofed audio~\cite{stan2025tada, mishra2026towards}, it often relies on an implicit assumption: that ``authentic" audio remains a single, pristine distribution. This assumption becomes especially fragile for partially processed audio, where an utterance may contain both untouched speech and locally modified regions.

High-fidelity media production relies heavily on signal processing chains, including modern speech enhancement and restoration~\cite{nakata2025sidon}. Moreover, stylistic modifications like voice quality conversion to creaky phonations -- while sometimes used adversarially to defeat neural audio watermarking~\cite{ozer2026self} -- can also be legitimately used to enhance paralinguistic expression~\cite{tsvetanova2017multimodal, zimman2017transgender}.

In this work we demonstrate how current spoof detection systems struggle under this benign authenticity processing, as benign processing artefacts are frequently misclassified as spoofing. Maintaining a rigid framework where any processed signal might be flagged as spoofed increases false-positive risk in practical deployments. In anti-spoofing deployments, the objective is typically to detect malicious impersonation rather than catch quality-of-life improvement artefacts added by the processing~\cite{zuo2024codecfake,yamagishi2021asvspoof}. 
The problem is not simply that processed bona fide speech may be misclassified, but that a single authentic/spoof label cannot specify which property the detector is evaluating: the source of the speech, the presence of processing, or the location of an edit.
\section{Problem Formulation and Related work}
We use \textit{benign, authenticity-preserving transformations} to refer to transformations that preserve linguistic content and speaker source while modifying aspects such as voice quality or restoration state. 
We do not claim that such transformations are harmless in every listening context; ``benign'' here denotes preservation of speaker source and linguistic content, not the absence of perceptual or pragmatic effect.


Prior work on partially spoofed or partially edited speech has established the problem of detecting, localising spoofed regions embedded within an otherwise genuine utterance~\cite{zhang2021initial, zhang2022partialspoof, zhang2024spoofdiarization, zhang2025partialedit}. That literature generally assumes that the edited region is itself the spoof: the local modification is what makes part of the signal fake. We study a complementary case in which the edit is not a spoofing operation at all, but a source-preserving transformation such as voice quality conversion (VQC) or restoration. Such benign changes edit speech without changing the speaker source or linguistic content, so an utterance-level authentic/spoofed label becomes under-specified: it no longer says enough whether a spoofing detector should report source authenticity, processing status, or the location of the edit.

We hypothesise that binary spoof detectors conflate source authenticity and processing status, and make the following contributions:
\begin{itemize}
    \item We reformulate the deepfake task as a 4-way task that exposes the difficulty of separating source authenticity from processing status.
    \item We show, across SSL embeddings and DF-Arena fine-tuning, that utterance processing status can be easier to detect than source authenticity when we test on out-of-domain samples.
    \item We introduce a partial-processing VQC stress test showing that low binary Equal Error Rates (EER) can hide poor detection of local processing.
    \item We release processed bona fide and spoofed speech data to support evaluation of anti-spoofing systems at this~\href{https://zenodo.org/records/18803182}{\blue{link}}.
\end{itemize}

\section{Dataset}
To evaluate how authenticity-preserving transformations affect spoofing detection, we build upon and create a dataset that pairs bona fide audio with spoofed counterparts along with their processed versions.

\subsection{Corpora and Synthetic speech (TTS) sources}
We used the paired utterances from the real M-AILABS corpus \cite{mailabs_2017} and the deepfakes from the MLAAD corpus~\cite{mlaad}. M-AILABS contains English audiobook recordings from LibriVox. The corpus includes both male and female speakers recorded in quiet conditions.  MLAAD \cite{mlaad} contains synthetic data that is synthesised using M-AILABS as its bona fide data source. We selected 2,575 utterances that have matching (Text-to-Speech) TTS counterparts across 10 TTS systems, ensuring balanced representation across speakers and content. All audio was resampled to 16\,kHz for consistency.

We use 10 diverse TTS architectures: FireRedTTS-2.0~\cite{zhou2025firered}, Higgs-Audio-V2~\cite{liu2025higgs}, Index-TTS-2.0~\cite{zhou2025index}, Llasa-1B~\cite{zuo2024llasa}, MiniCPM-o-2.6~\cite{openbmb2025minicpm}, Openaudio-S1-Mini~\cite{fishaudio2025openaudio}, OuteTTS~\cite{outeai2024outetts}, VoxCPM-0.5B~\cite{zhou2025voxcpm}, VoXtream~\cite{torgashov2025voxtream}, and ZipVoice~\cite{zhu2025zipvoice}. These architectures were selected because they were annotated with reference speaker information. This allowed us to select only the utterances for which the reference speaker was identical to the source speaker from the original M-AILABS corpus. Each TTS system synthesises the same text content as the bona fide recordings, yielding 2,575 matched utterance pairs. This ensures that any observed differences between bona fide and spoofed embeddings are attributable to source characteristics rather than linguistic content or speaker conversion.

\subsection{Benign Speech Transformations}

We evaluate our hypotheses using two benign transformation types: (1) Voice Quality Conversion (VQC) and (2) speech restoration. Unlike generative attacks designed for impersonation, these transformations represent intra-speaker variations that a robust spoofing countermeasure should not treat as synthetic source evidence. The models employed here report high speaker similarity metrics pre and post-transformation~\cite{nakata2025sidon, li2023freevc}. In the case of VQC, the glottal source parameters are modified while preserving the semantic content and minimising speaker identity changes. This allows for enhanced paralinguistic and pragmatic expression, for instance, using breathy voice to signal intimacy~\cite{tsvetanova2017multimodal} or creaky voice to indicate a turn yield~\cite{heldner2019voice}. From the available VQC frameworks~\cite{rautenberg, lameris2024creakvc}, we chose~\cite{lameris25_interspeech} due to its support for multiple phonation types. We converted 2,575 utterance pairs into four categories: modal, breathy, creaky, and end-creak. These types were selected because they account for the most common phonation types in English~\cite{podesva2011gender} and serve diverse pragmatic functions, such as marking parenthetical comments~\cite{lee2015creaky} or signalling utterance termination~\cite{henton1989sociophonetic}. An additional transformation type is speech restoration. Speech restoration models, such as Sidon~\cite{nakata2025sidon}, generate restored speech using representations from speech foundation models. Across both transformed scenarios, we use the following labels:

\begin{itemize}
    \item \textbf{Bona fide:} Unprocessed recordings of human speech.
    \item \textbf{Spoofed:} Unprocessed synthetic speech generated via TTS.
    \item \textbf{Processed Bona fide:} Bona fide speech after benign processing (VQC or speech restoration).
    \item \textbf{Processed Spoofed:} Spoofed speech after the same benign processing.
\end{itemize}
\begin{figure}[!t]
    \centering
    \includegraphics[width=\linewidth]{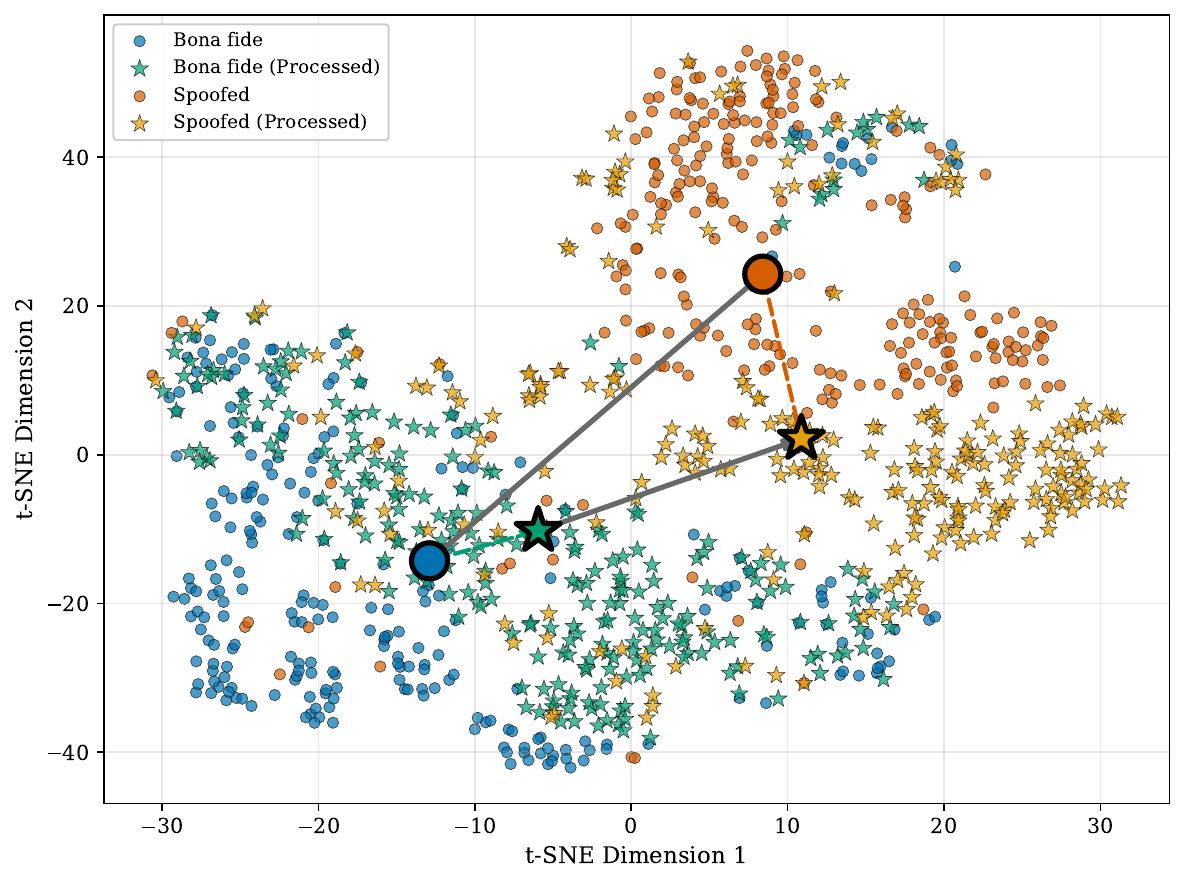}
    \caption{t-SNE plots of Wav2Vec2 embeddings before and after Sidon enhancement on MLAAD matched dataset.}
    \label{fig:wav2vec2_sidon}
\end{figure}

For out-of-domain evaluations on the ASVspoof5 dataset, we report results on VQC class-balanced subsets (2,000 utterances per class; 8,000 total) sampled without replacement. For mixed-domain fine-tuning, we construct a disjoint ASVspoof5 train/val/test partition by first sampling an equal number of utterances per class and then reserving a held-out test set of 2,000 per class; the remaining samples are split into train/validation (approximately 70/15). The ASVspoof5 test partition is never used for training, early stopping, or model selection. Further out-of-domain evaluations with Sidon are used to restore the same 2,575 MLAAD utterance pairs, resulting in bona fide enhanced/restored audio and spoofed enhanced/restored audio. 

We reported our results as both mean and standard deviations over five random seeds, where each seed controls model initialisation, data-loader ordering, and fine-tuning data sampling/splitting. Our processed datasets are available~\href{https://zenodo.org/records/18803182}{\blue{here}}.
\section{Methodology and Experiments}
We first use SSL embedding visualisations and directional consistency to characterise how the transformations move bona fide and spoofed speech in representation space. We then present 4-way classification results, a partial-processing stress test, and an acoustic analysis involving glottal source parameters and spectral tilt which offers interpretations to the VQC classification results.

\subsection{Embedding analysis}
\begin{figure}[!t]
    \centering
    \includegraphics[width=\linewidth]{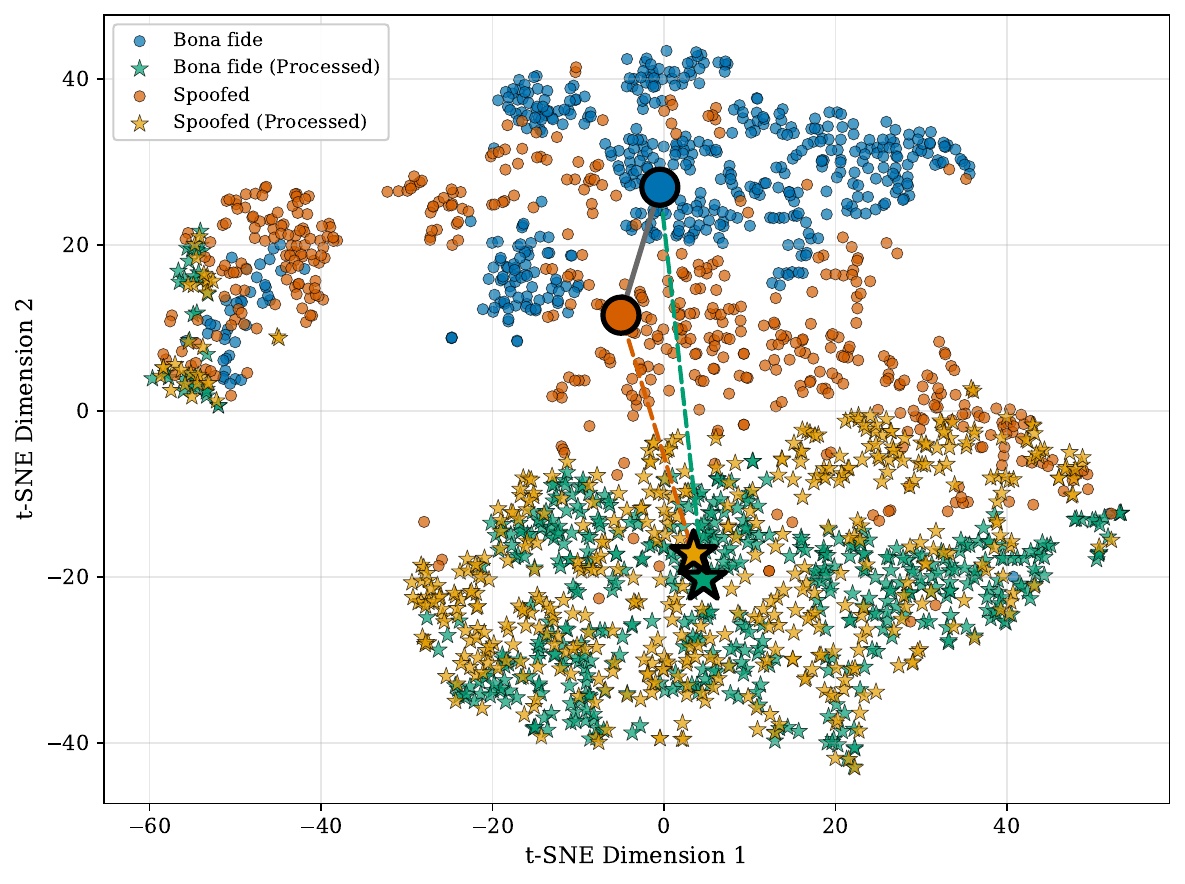}
    \caption{t-SNE plots of Wav2Vec2 embeddings before and after VQC on MLAAD matched dataset.}
    \label{fig:wav2vec2_vqvc}
\end{figure}

We use mean-pooled SSL embeddings from HuBERT-base~\cite{HuBERT2021}, Whisper-small~\cite{Whisper2023}, and Wav2Vec2 XLS-R 1B~\cite{XLSR2022} to analyse how our chosen transformations alter the representation space used for spoof detection. To test whether VQC induces similar shifts regardless of source type, we compute \emph{directional consistency} between bona fide and spoofed shift vectors.
For each source type $s \in \{\text{bona fide}, \text{spoofed}\}$, we compute the mean embedding shift vector:
\begin{equation}
    \Delta_s = \frac{1}{N_s} \sum_{x \in s} \left( \Emb(\vqc(x)) - \Emb(x) \right).
\end{equation}
Directional consistency is then measured as the cosine similarity between these mean shift vectors:
\begin{equation}
    \cos(\Delta_{\mathrm{bonafide}}, \Delta_{\mathrm{spoofed}}) = \frac{\Delta_{\mathrm{bonafide}} \cdot \Delta_{\mathrm{spoofed}}}{\lVert\Delta_{\mathrm{bonafide}}\rVert \lVert\Delta_{\mathrm{spoofed}}\rVert}
\end{equation}
A value near $+1$ indicates that VQC pushes bona fide and TTS embeddings in the same direction; a value near $0$ indicates orthogonal shifts; a negative value indicates opposing directions. 

\begin{figure}[t]
\centering
\includegraphics[width=\linewidth]{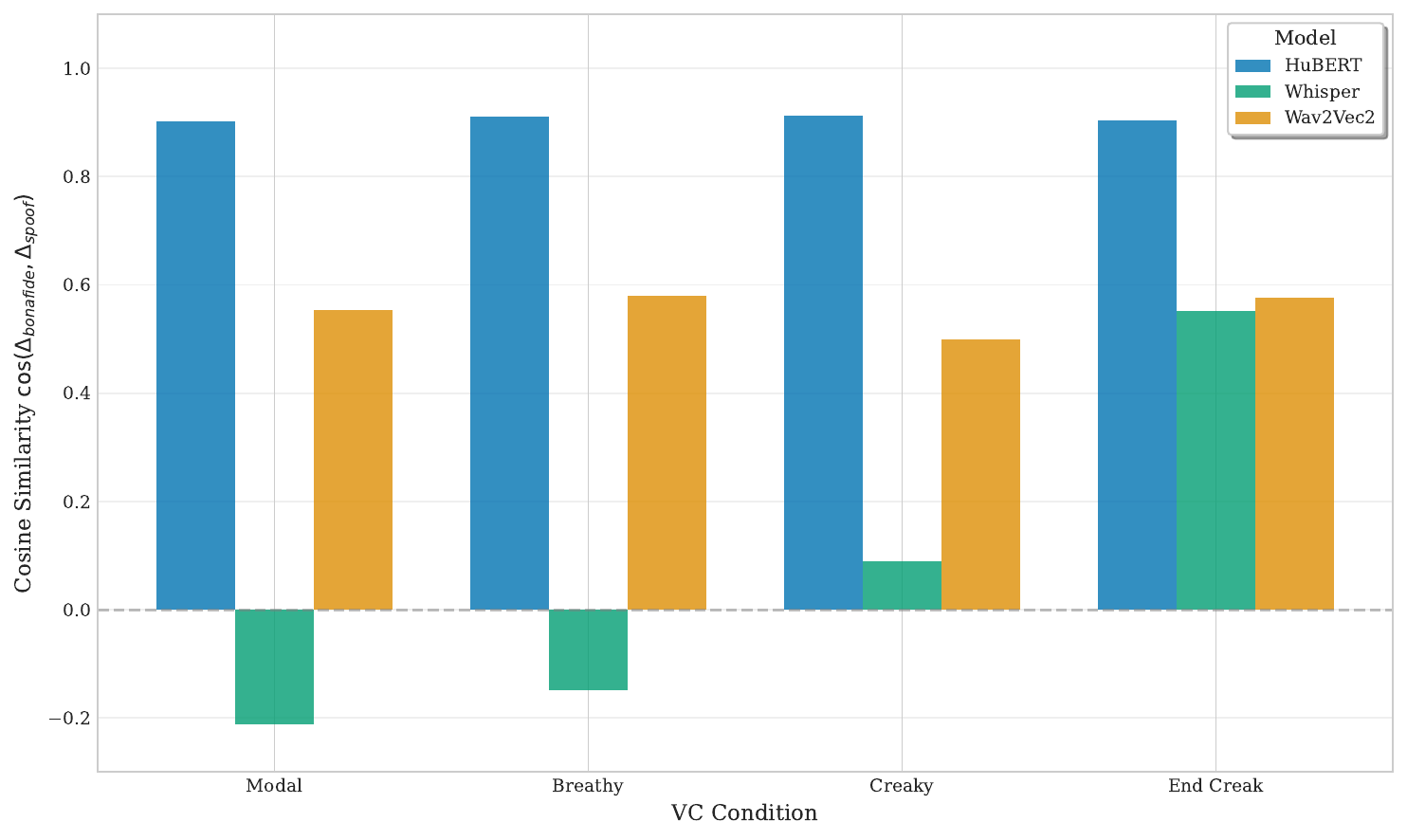}
\caption{Directional consistency of VQC-induced embedding shifts between mean bona fide and spoofed shift vectors. Whisper shows lower or negative values, suggesting potential source-dependent shift directions.}
\label{fig:direction}
\end{figure}

We first plot t-SNE embeddings of the Wav2Vec2 features, used in the DF-Arena classifiers, before and after the authenticity-preserving transformations in Figure~\ref{fig:wav2vec2_sidon} and Figure~\ref{fig:wav2vec2_vqvc} which show how the centroids of spoofed and bona fide Wav2Vec2 embeddings drift toward each other. We also find that the Whisper features have variable directional consistency under VQC as seen in Figure~\ref{fig:direction}. HuBERT and Wav2Vec2 show high consistency across all VQC conditions, indicating VQC applies a similar transformation to both source types. However, cosine similarity does not account for the magnitude of shifts and the source/speaker-dependent shift directions which may average out when aggregated.  

\subsection{Experiments}

We compare two classification architectures: An MLP trained on concatenated mean-pooled SSL embeddings from Wav2Vec2, HuBERT, and Whisper (2816 Dimensions) in binary and 4-way configurations and the pretrained DF-Arena\,1B~\cite{dowerah2026speech}, an open-source state-of-the-art binary anti-spoofing model built on a Wav2Vec2 XLS-R 1B backbone.

\noindent\textbf{Fine-tuning DF-Arena for multi-class detection:}
We replace DF-Arena's binary head (\texttt{fc5}: $1280{\to}2$) with a 4-class head ($1280{\to}4$), initialising classes~0 (bona fide) and~2 (spoofed) from the pretrained weights and classes~1 and~3 (processed variants of bona fide and spoofed) from the spoof weights since the pretrained model already maps all processed audio near its spoof representation. The Wav2Vec2 backbone is frozen; our default setting fine-tunes the final Conformer block and the classification head ($AdamW$, ${lr} = 10^{-4}$, 20 epochs). A binary variant uses the same protocol with a 2-class head. We select the checkpoint with the lowest validation loss and report results. 

Our initial experiments reveal that while models fine-tuned on MLAAD achieved ${\sim}99\%$ in-domain accuracy, they fail catastrophically out-of-domain: bona fide accuracy (${Acc}_{\mathrm{bona}}$) on ASVspoof5 drops to 0.1\%, as unseen bona fide speech collapses into the spoofed class. To recover cross-domain source detection, we continue from the MLAAD-only checkpoint with mixed-domain training, combining MLAAD and balanced ASVspoof5 data at a reduced learning rate ($5{\times}10^{-5}$) with early stopping ($patience = 8$). Since speech restoration induces a distributional shift distinct from VQC, we further augment the training set with Sidon-enhanced utterances, yielding the final models reported in Section~\ref{sec:results}. Because accuracy depends on the decision threshold (argmax for 4-way; fixed threshold for binary), we interpret performance primarily through threshold-free metrics ($\mathrm{EER}_{\mathrm{src}}$ and, for 4-way models, $\mathrm{EER}_{\mathrm{proc}}$). 

\section{Results}\label{sec:results}

\begin{table*}[t]
\centering
\caption{Summary of classification results. DF-Arena rows report mean $\pm$ standard deviation over five random seeds. MLAAD metrics use full in-domain test sets (Seen = 9 TTS architectures; Unseen = held-out OuteTTS). ASVspoof5 is evaluated on held-out class-balanced splits. Rows trained with Sidon use held-out Sidon test-split utterances. For 4-way models, $\mathrm{EER}_{\mathrm{proc}}$ collapses predictions along the \textit{processed} axis. $\mathrm{Acc}_{\mathrm{bona}}$ measures accuracy on bona fide-source speech.}
\label{tab:all-experiments}
\smallskip
\setlength{\tabcolsep}{1.5pt}
\renewcommand{\arraystretch}{1.05}
\resizebox{\textwidth}{!}{%
\begin{tabular}{@{}llc|rrrr|rrrr|rrr@{}}
\toprule
 & & & \multicolumn{4}{c|}{\textbf{MLAAD VQC}} & \multicolumn{4}{c|}{\textbf{ASVspoof5 VQC}} & \multicolumn{3}{c}{\textbf{MLAAD Sidon}} \\
\cmidrule(lr){4-7} \cmidrule(lr){8-11} \cmidrule(lr){12-14}
\textbf{Model} & \textbf{Training/Fine-tuning data} & \textbf{\#\,Cls} & \multicolumn{1}{c}{Seen Acc} & \multicolumn{1}{c}{Seen EER} & \multicolumn{1}{c}{Unseen Acc} & \multicolumn{1}{c|}{Unseen EER} & \multicolumn{1}{c}{Acc} & \multicolumn{1}{c}{$\mathrm{EER}_{\mathrm{src}}$} & \multicolumn{1}{c}{$\mathrm{EER}_{\mathrm{proc}}$} & \multicolumn{1}{c|}{$\mathrm{Acc}_{\mathrm{bona}}$} & \multicolumn{1}{c}{Acc} & \multicolumn{1}{c}{$\mathrm{EER}_{\mathrm{src}}$} & \multicolumn{1}{c}{$\mathrm{Acc}_{\mathrm{bona}}$} \\
\midrule
\multicolumn{14}{@{}l}{\textit{SSL-embedding MLP; HuBERT $\oplus$ Whisper $\oplus$ Wav2Vec2}} \\[3pt]
\quad Binary & MLAAD VQC & 2 & 99.1 & 0.81 & 90.2 & 4.18 & 48.0 & 54.7 & -- & 6.2 & 88.6 & 5.62 & 78.8 \\
\quad 4-Way & MLAAD VQC & 4 & 99.0 & 0.88 & 89.6 & 4.38 & 43.5 & 52.3 & 2.4 & 24.0 & 76.7 & 10.01 & 55.0 \\
\midrule
\multicolumn{14}{@{}l}{\textit{DF-Arena 1B fine-tuned; Wav2Vec2, five seeds}} \\[3pt]
\quad Binary fine-tuned & MLAAD VQC & 2 & $98.99{\pm}0.07$ & $1.02{\pm}0.17$ & $97.63{\pm}0.28$ & $2.10{\pm}0.29$ & $49.92{\pm}0.02$ & $56.13{\pm}6.89$ & -- & $0.07{\pm}0.05$ & $56.27{\pm}1.55$ & $17.08{\pm}8.98$ & $11.79{\pm}3.13$ \\
\quad 4-Way fine-tuned & MLAAD VQC & 4 & $98.85{\pm}0.15$ & $1.14{\pm}0.12$ & $97.88{\pm}0.19$ & $2.22{\pm}0.30$ & $49.58{\pm}0.29$ & $49.52{\pm}4.59$ & $0.20{\pm}0.07$ & $0.10{\pm}0.06$ & $1.46{\pm}0.92$ & $14.73{\pm}3.00$ & $14.41{\pm}3.50$ \\
\quad Binary fine-tuned & (MLAAD + ASVspoof5) VQC & 2 & $98.29{\pm}0.31$ & $1.53{\pm}0.35$ & $97.30{\pm}0.59$ & $2.30{\pm}0.57$ & $84.75{\pm}2.48$ & $12.90{\pm}1.85$ & -- & $75.05{\pm}5.39$ & $56.26{\pm}2.71$ & $38.20{\pm}10.14$ & $11.80{\pm}5.53$ \\
\quad 4-Way fine-tuned & (MLAAD + ASVspoof5) VQC & 4 & $97.68{\pm}0.73$ & $1.93{\pm}0.32$ & $96.91{\pm}0.60$ & $2.57{\pm}0.44$ & $85.80{\pm}2.30$ & $13.08{\pm}1.66$ & $0.01{\pm}0.01$ & $81.12{\pm}6.72$ & $0.67{\pm}0.60$ & $37.77{\pm}6.38$ & $8.62{\pm}2.20$ \\
\quad Binary fine-tuned$^*$ &  MLAAD (VQC + Sidon) + ASVspoof5 VQC & 2 & $98.11{\pm}0.62$ & $1.57{\pm}0.47$ & $96.87{\pm}1.18$ & $2.33{\pm}0.31$ & $86.72{\pm}1.02$ & $11.93{\pm}0.65$ & -- & $80.78{\pm}4.56$ & $93.62{\pm}3.77$ & $2.27{\pm}0.39$ & $88.05{\pm}8.00$ \\
\quad 4-Way fine-tuned$^*$ & MLAAD (VQC + Sidon) + ASVspoof5 VQC  & 4 & $98.19{\pm}0.18$ & $1.68{\pm}0.17$ & $97.22{\pm}0.64$ & $2.37{\pm}0.44$ & $86.48{\pm}0.78$ & $11.06{\pm}1.17$ & $0.14{\pm}0.11$ & $78.41{\pm}1.80$ & $90.23{\pm}4.69$ & $2.60{\pm}1.05$ & $84.51{\pm}10.14$ \\
\bottomrule
\end{tabular}%
}
\end{table*}

Table~\ref{tab:all-experiments} summarises in-domain evaluation on MLAAD VQC (Seen/Unseen TTS), out-of-domain evaluation on ASVspoof5, and Sidon restoration. For 4-way models, $\mathrm{EER}_{\mathrm{proc}}$ collapses predictions along the processing axis (unprocessed vs.\ processed), while $\mathrm{EER}_{\mathrm{src}}$ collapses predictions along the source-authenticity axis (bona fide vs.\ spoofed). This separates the conflation that is inherent to the binary anti-spoofing formulation: whether speech was processed, and whether its source is synthetic or not.

The first pattern is that binary models often treat processing evidence as evidence of spoofing. MLAAD-only fine-tuning gives high in-domain accuracy, but it plummets under cross-corpus evaluation: ASVspoof5 bona fide-source accuracy falls to almost zero for both the binary and 4-way MLAAD-only DF-Arena variants. This indicates that the models do not learn a notion of source authenticity; instead, they rely on cues tied to the training distribution and processing status.

The second pattern is that full-utterance processing status transfers more reliably than source attribution. With mixed-domain training, the 4-way DF-Arena model reaches very low $\mathrm{EER}_{\mathrm{proc}}$ on ASVspoof5, while $\mathrm{EER}_{\mathrm{src}}$ remains much larger. The value of our 4-way class formulation is diagnostic: it exposes cases where a detector separates processing status while still failing to separate source authenticity. 

\begin{table}[!b]
\centering
\caption{Single seed Confusion matrix of mixed-domain DF-Arena 4-way on ASVspoof5. B = Bona fide, S = Spoofed, B$\to$P = Processed Bona fide, and S$\to$P = Processed Spoofed.}
\label{tab:cm_asvspoof5_mixed}
\footnotesize
\setlength{\tabcolsep}{5pt}
\renewcommand{\arraystretch}{1.1}
\begin{tabular}{lrrrr}
\toprule
\diagbox[width=6em]{True}{Pred.} & \textbf{B} & \textbf{B$\to$P} & \textbf{S} & \textbf{S$\to$P} \\
\midrule
\textbf{B}      & 1894 &    0 &  106 &    0 \\
\textbf{B$\to$P} &    0 & 1317 &    0 &  683 \\
\textbf{S}      &   38 &    0 & 1960 &    2 \\
\textbf{S$\to$P} &    0 &  229 &    0 & 1771 \\
\bottomrule
\end{tabular}
\end{table}

Table~\ref{tab:cm_asvspoof5_mixed} shows the unresolved failure mode directly. The model separates unprocessed from processed speech almost perfectly, but source attribution within processed speech is much less reliable: 683 processed bona fide samples are assigned to processed spoofed. Mixed-domain exposure improves source performance, but the remaining confusions show why source authentication and processing detection should not be reduced to one utterance-level label.

Benign transformations do not share a single transferable ``signature'': Fine-tuning with VQC exposure does not make the binary model robust to Sidon-restored bona fide speech ($11.80{\pm}5.53\%$ bona fide accuracy). Adding Sidon-augmented train-split data recovers Sidon performance for both binary and 4-way DF-Arena models, with binary higher on several Sidon accuracy metrics.


\subsection{Partial-Processing Stress Test}

We do not consider end-creak setting to be conventional partial processing since the whole utterance is resynthesised during voice quality conversion. To connect the utterance-level experiments to partially edited setting, we construct partial VQC processing by replacing only a fixed proportion of contiguous speech-active regions with VQC-converted counterparts. This creates utterances whose source authenticity is unchanged but whose processed proportion varies from 10\% to 50\%. We evaluate trained checkpoints on held-out partial VQC mixtures that were not used during training. The models have seen full-utterance VQC, but not localised VQC at controlled processed ratios. We use the five-seed mixed-domain binary and 4-way DF-Arena checkpoints and evaluate on both seen and unseen TTS splits. 
\begin{table}[!b]
\centering
\caption{Binary paired perturbation proxies for bona fide-source partial VQC. The gold binary label remains bona fide; metrics compare each partially processed utterance with its unprocessed original. Values are mean$\pm$std over five seeds.}
\label{tab:binary_partial_proxy}
\small
\setlength{\tabcolsep}{3pt}
\resizebox{\columnwidth}{!}{%
\begin{tabular}{lrrrr}
\toprule
\textbf{Split/Proc.} & $\Delta p_{\mathrm{spoof}}$ (pp) & $P(\Delta p_{\mathrm{spoof}}>5\mathrm{pp})$ (\%) & $P(\Delta\mathrm{conf.}>5\mathrm{pp})$ (\%) & \textbf{Flip} (\%) \\
\midrule
Seen 10\% & $0.27{\pm}0.55$ & $1.19{\pm}2.33$ & $1.11{\pm}2.17$ & $0.12{\pm}0.26$ \\
Seen 25\% & $1.89{\pm}2.18$ & $4.62{\pm}5.66$ & $4.19{\pm}5.19$ & $1.55{\pm}1.90$ \\
Seen 50\% & $8.76{\pm}7.19$ & $17.74{\pm}13.01$ & $15.15{\pm}10.97$ & $7.87{\pm}7.01$ \\
\midrule
Unseen 10\% & $0.29{\pm}0.47$ & $1.09{\pm}2.23$ & $1.09{\pm}2.23$ & $0.16{\pm}0.21$ \\
Unseen 25\% & $1.66{\pm}2.32$ & $4.36{\pm}5.76$ & $4.20{\pm}5.44$ & $1.32{\pm}2.12$ \\
Unseen 50\% & $8.65{\pm}7.72$ & $18.52{\pm}13.08$ & $16.34{\pm}10.89$ & $7.63{\pm}7.65$ \\
\bottomrule
\end{tabular}%
}
\end{table}

Table~\ref{tab:binary_partial_proxy} shows the results of partial VQC processing. We measure \begin{enumerate}
\item \textbf{$\Delta$ p\_spoof (pp)}: average increase in the binary model’s spoof probability, in percentage points, after partial VQC.
\item \textbf{P($\Delta$ p\_spoof $>$ 5pp)}: percent of bona fide samples where spoof probability increases by more than 5 points.
\item \textbf{P($\Delta$ conf. $>$ 5pp)}: percent where the model’s confidence drops by more than 5 points.

\end{enumerate}
These measures show that benign local processing perturbs binary spoof confidence asymmetrically, especially for bona fide speech, but binary outputs still cannot say whether processing occurred or where it occurred.

Table~\ref{tab:partial_processing} shows why source-only evaluation is incomplete for partially processed speech. Under the binary source label, partial VQC remains bona fide for human speech and spoofed for TTS speech. With that label definition, the task appears almost solved: even at 50\% processed ratio, source EER remains below 4\% across seen and unseen splits. However, low source EER does not mean that processing was detected. The 4-way processing EER is close to chance for 10\% processed spans ($47.17{\pm}1.55\%$ seen; $48.17{\pm}1.58\%$ unseen), improves at 25\%, and remains high at 50\% ($27.99{\pm}1.63\%$ seen; $27.20{\pm}2.69\%$ unseen). Thus, a source-authentication metric can make partial VQC look solved while the local processing event remains weakly exposed by utterance-level processing scores.

\begin{table}[t]
\centering
\caption{Partial-processing VQC stress test. Results are EER\% mean$\pm$std over five seeds.}
\label{tab:partial_processing}
\small
\setlength{\tabcolsep}{3pt}
\resizebox{\columnwidth}{!}{%
\begin{tabular}{lrrr}
\toprule
\textbf{Split/Proc.} & \textbf{Binary} $\mathrm{EER}_{\mathrm{src}}$ & \textbf{4-way} $\mathrm{EER}_{\mathrm{src}}$ & \textbf{4-way} $\mathrm{EER}_{\mathrm{proc}}$ \\
\midrule
Seen 10\% & $0.04{\pm}0.10$ & $0.01{\pm}0.03$ & $47.17{\pm}1.55$ \\
Seen 25\% & $0.31{\pm}0.18$ & $0.25{\pm}0.13$ & $42.52{\pm}1.90$ \\
Seen 50\% & $1.83{\pm}1.39$ & $3.38{\pm}1.32$ & $27.99{\pm}1.63$ \\
\midrule
Unseen 10\% & $0.00{\pm}0.00$ & $0.00{\pm}0.00$ & $48.17{\pm}1.58$ \\
Unseen 25\% & $0.08{\pm}0.17$ & $0.23{\pm}0.21$ & $42.53{\pm}3.29$ \\
Unseen 50\% & $1.17{\pm}1.65$ & $2.57{\pm}1.42$ & $27.20{\pm}2.69$ \\
\bottomrule
\end{tabular}%
}
\end{table}

\subsection{Acoustic Analysis}

\begin{figure*}[!t]
    \centering
    \includegraphics[width=0.9\linewidth]{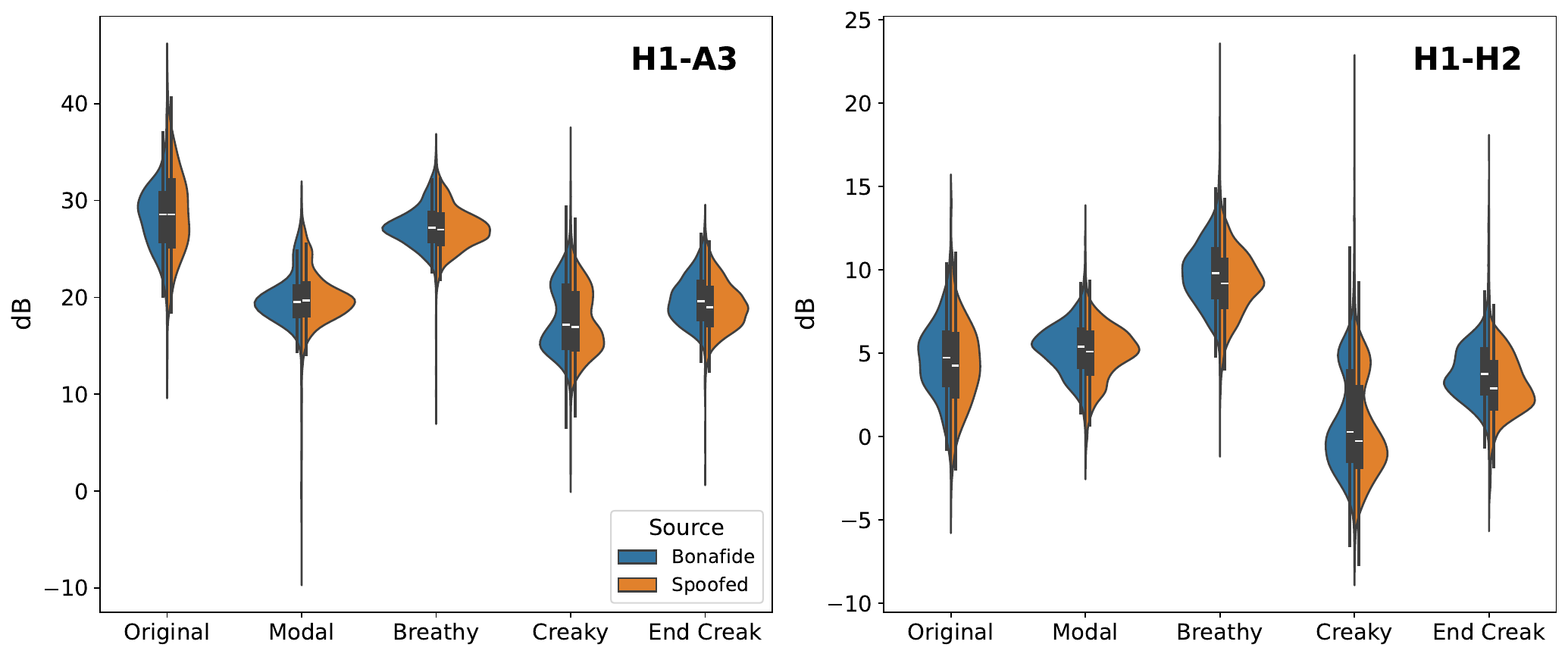}
    \caption{The acoustic feature shifts between the original M-AILABS/MLAAD recordings and the converted recordings by voice quality.}
    \label{fig:acoustic_analysis_fig}
\end{figure*}

In order to investigate the glottal source parameters and the acoustic shifts by the transformations, we measured H1--A3, a spectral measure related to the abruptness of closure of the vocal folds and H1--H2, a spectral measure related to the open quotient, i.e. the fraction of time that the vocal folds allow the passage of air on the bona fide (M-AILABS) and spoofed (MLAAD) recordings, as well as the converted and enhanced versions. Results can be found in Figure~\ref{fig:acoustic_analysis_fig}. The measurements were analysed with a two-way ANOVA with Tukey HSD for each spectral measure with the main effects of source (bona fide or spoofed) and processing type (target voice quality), as well as the interaction. For both H1--A3 and H1--H2, there was a strongly significant main effect of source ($p < 0.0001$) as well as processing type ($p < 0.0001$). Additionally, a highly significant interaction effect was observed for both measures ($p < 0.0001$), demonstrating that VQC non-uniformly increases the acoustic differences. For the original bona fide and spoofed recordings, the sources exhibited no statistically significant differences in H1--A3 ($p = 0.7403$) or H1--H2 ($p = 0.0548$). Nevertheless, the VQC conversion widens the gap between spoofed and authentic samples and reaches its maximum in the end creak condition (interaction deltas of $-0.99$ dB for H1--A3 and $-0.61$ dB for H1--H2, $p < 0.0001$). 
The acoustic analysis of H1--A3 and H1--H2 for Sidon restored audio yielded significant main effects for source authenticity and processing type, but no interaction effect, indicating that the enhancement process induced a consistent global shift for both bona fide and spoofed samples.
\begin{table}[!b]
\centering
\caption{The acoustic differences of the voice quality transformations between spoofed and bona fide audio}
\label{tab:acoustic_analysis}
\begin{tabular}{l c c}
\hline
\textbf{Spoofed -- Bona fide} & \textbf{H1--A3} & \textbf{H1--H2} \\
\hline
{MLAAD Unconverted Gap} & $+0.36$ dB & $-0.38$ dB \\
 & $p = .7403$ & $p = .0548$ \\
\hline
{Modal Int.} ($\Delta$) & $-0.02$ dB & $-0.01$ dB \\
{Breathy Int.} ($\Delta$) & $-0.54$ dB & $-0.36$ dB \\
{Creaky Int.} ($\Delta$) & $-0.81$ dB & $-0.33$ dB \\
{End Creak Int.} ($\Delta$) & $-0.99$ dB & $-0.61$ dB \\
\hline
\textbf{Interaction} ($p$-value) & $p < .0001$ & $p < .0001$ \\
\hline
\end{tabular}
\end{table}
These findings confirm that while the original recordings are acoustically similar, VQC exaggerates latent synthetic artefacts. While Sidon restoration does not interact with voice quality features, our analysis offers interpretable explanations as to why classifiers struggle with distinguishing out-of-domain processing and offers discriminatory features for improved detection of transformation source.
\section{Discussion}

\noindent\textbf{Binary labels are under-specified for partial processing.}
A single bona fide/spoofed label is often insufficient when an utterance contains both untouched and processed regions. In such cases, the source of the speech and the presence of processing are distinct properties. A bona fide speaker may produce an utterance that is later restored, enhanced, or locally modified without becoming a synthetic impersonation. Conversely, spoofed speech may also be processed after generation. Treating these cases as a single binary decision encourages detectors to use processing artefacts as a proxy for spoofing.

\noindent\textbf{What the 4-way formulation does and does not solve.}
The proposed 4-way classification is not intended as a final taxonomy of all possible transformations. Our results show that full-utterance processing status can generalise strongly across domains, while source attribution under processing remains fragile. This explains why 4-way supervision should not be interpreted simply as a robustness improvement over binary training; its main value is making the failure mode measurable. The core classes are source authenticity $\times$ processing status, while specific transformation types such as VQC or restoration are additional information.

\noindent\textbf{Toward region-aware defences.}
The partial-processing experiment suggests that source accuracy alone can give false reassurances: a system may correctly preserve the bona fide/spoofed source decision while failing to expose local processing. The processing signal itself depends strongly on processing extent and is weak when only a short span is modified. This is important for real deployment, where short processed spans may be embedded inside otherwise genuine recordings. Future defences should therefore combine utterance-level source attribution with localisation or diarisation-style estimates of where processing occurs. Such systems would allow downstream users to distinguish fully synthetic speech, locally processed bona fide speech, and processed synthetic speech.

\section{Conclusion}
We showed that binary deepfake detectors can conflate source authenticity with benign processing/edits. Voice quality conversion, speech restoration, and partial processing introduce cues that are often easier to detect than the underlying bona fide/spoofed source when applied to full utterances, but harder to detect when localised to short regions. We reformulated the deepfake detection task as a 4-way classification to expose this separation. The results motivate transformation-aware evaluation protocols for partially processed audio, where systems report not only whether speech is spoofed, but also whether it was processed and where that processing occurs.




\bibliographystyle{IEEEtran}
\bibliography{mybib}

\end{document}